\documentclass[%
reprint,
superscriptaddress,
 amsmath,amssymb,
 aps,
prb,
]{revtex4-1}

\usepackage {graphicx,epsfig,graphics,color}
\usepackage{dcolumn}
\usepackage{bm}
\usepackage{hyperref}
\usepackage{sidecap}
\usepackage{float}

\begin{document}
\title{Thermally populated versus field-induced triplon bound states in the Shastry-Sutherland lattice SrCu$_2$(BO$_3$)$_2$}

\author{Dirk Wulferding}
\email[]{Corresponding author: dirwulfe@snu.ac.kr}
\altaffiliation{Contributed equally to this work.}
\affiliation{Center for Correlated Electron Systems, Institute for Basic Science, Seoul 08826, Republic of Korea}
\affiliation{Department of Physics and Astronomy, Seoul National University, Seoul 08826, Republic of Korea}

\author{Youngsu Choi}
\altaffiliation{Contributed equally to this work.}
\affiliation{Department of Physics, Chung-Ang University, Seoul 06974, Republic of Korea}

\author{Seungyeol Lee}
\affiliation{Department of Physics, Chung-Ang University, Seoul 06974, Republic of Korea}

\author{Mikhail A. Prosnikov}
\affiliation{High Field Magnet Laboratory (HFML-EMFL), Radboud University, 6525 ED Nijmegen, Netherlands}

\author{Yann Gallais}
\affiliation{Laboratoire Mat\'{e}riaux et Ph\'{e}nom\`{e}nes Quantiques (UMR 7162 CNRS), Universit\'{e} de Paris, 75205 Paris Cedex 13, France}

\author{Peter Lemmens}
\affiliation{Institute for Condensed Matter Physics, TU Braunschweig, D-38106 Braunschweig, Germany}

\author{Chengchao Zhong}
\affiliation{Department of Energy and Hydrocarbon Chemistry, Graduate School of Engineering, Kyoto University, Kyoto 615-8510, Japan}

\author{Hiroshi Kageyama}
\affiliation{Department of Energy and Hydrocarbon Chemistry, Graduate School of Engineering, Kyoto University, Kyoto 615-8510, Japan}

\author{Kwang-Yong Choi}
\email[]{Corresponding author: choisky99@skku.edu}
\affiliation{Department of Physics, Sungkyunkwan University, Suwon 16419, Republic of Korea}

\date{\today}

\begin{abstract}
The Shastry-Sutherland compound SrCu$_2$(BO$_3$)$_2$ constituting orthogonally coupled dimers harbors a $S=0$ singlet ground state. The confluence of strong interdimer interaction and frustration engenders a spectrum of low-energy excitations including localized triplons as well as singlet and triplet bound states. Their dynamics are controlled by an external magnetic field and temperature. Here, we employ high-field Raman spectroscopy to map the field and temperature evolution of such bosonic composite quasiparticles on approaching the 1/8 magnetization plateau. Our study unveils that the magnetic field and thermal fluctuations show remarkably similar effects in melting the singlet bound states, but are disparate in their effects on the fine spectral shapes. This, together with the anti-crossing of two singlet bound states in the intermediate field $B=10-16$~T, is discussed in terms of the correlated dynamics of frustrated, interacting bosons.



\end{abstract}

\maketitle

\section{Introduction}

Among many-body quantum magnets, the 2D Shastry-Sutherland lattice (SSL) consisting of orthogonally coupled dimers inhabits a special place due to its rich physics resulting from a competition of intradimer and interdimer interactions $J$ and $J'$, respectively (see Fig. 1a)~\cite{shastry-81}. It is well established that the SSL has an exact dimer state at $J' \lesssim 0.675J$  and a gapless antiferromagnetic order at $J' \gtrsim 0.9J$~\cite{miyahara-03}. Remarkably, a narrow quantum spin-liquid phase at $J'\approx 0.77J$ has been proposed in addition to an intermediate plaquette singlet phase for $0.675J \lesssim J' \lesssim 0.76J$~\cite{koga-00,corboz-13,Zayed-17,Lee-19,wietek-19,guo-20,yang-21}. Examples of exotic physics that can be explored within this model include topologically non-trivial magnon bands with protected chiral triplon edge modes~\cite{romhanyi-15, mcclarty-17}, triplon Hall effect~\cite{malki-17}, a discontinuous Ising quantum critical point~\cite{jimenez-21}, a crystallization of spin superlattices as a function of pressure and magnetic field~\cite{haravifard-16, shi-21}, and a crossover to a quantum spin liquid phase~\cite{yang-21}.

One of the closest experimental realizations of a 2D SSL is SrCu$_2$(BO$_3$)$_2$~\cite{kageyama-99,miyahara-99}. Here, Cu$^{2+}$ ions form layers of $S=1/2$ spins alternatingly coupled via superexchange interactions $J$ and $J'$. The magnetic behavior of this SSL compound is described by the spin Hamiltonian~\cite{shastry-81}
\begin{equation}
\mathcal{H} = J \sum\limits_{nn} S_i \cdot S_j + J' \sum\limits_{nnn} S_i \cdot S_j - g \mu_B H \sum\limits_{i} S_i.
\end{equation}
With the ratio $J'/J \approx 0.63$~\cite{knetter-00}, SrCu$_2$(BO$_3$)$_2$ stabilizes a network of orthogonal spin dimers with a gap $\Delta/k_\mathrm{B}=35$~K as shown in Fig. 1b, albeit in close vicinity to a quantum critical point that separates the singlet ground state of orthogonal dimers from a plaquette singlet state~\cite{miyahara-03}.

Applying a magnetic field populates this singlet vacuum state by promoting triplet states (``triplons''), thereby enabling a control over the triplon density. Due to competing magnetic exchange interactions beyond nearest-neighbor coupling, the hopping of triplons is inhibited and they are extremely localized. This results in very flat triplon bands, as observed, e.g., via inelastic neutron scattering experiments~\cite{kageyama-00}, as well as multiple magnetization plateaus~\cite{onizuka-00,matsuda-13}.

At intermediate magnetic fields below saturation, the existence of these plateaus suggests that the suppressed kinetic energy of the triplons is conducive to a Wigner crystallization of triplon bound states at fractional fillings. This allows for tuning superlattice structures of bound states via applied magnetic field strength~\cite{momoi-00}. Spectroscopically, this was first observed at the 1/8 plateau via nuclear magnetic resonance~\cite{kodama-02,takigawa-10}. Simulations based on the infinite projected entangled-pair states method (iPEPS) revealed that at the 1/8 magnetization plateau bound states of triplets crystallize into a pinwheel-type pattern with unit cell vectors (4,2) and (0,4) within the $ab$ plane (see the simplified sketch in Fig. 1c and Ref. [\onlinecite{corboz-14}] for details). While bulk magnetization gives clues about the emergence of triplet patterns in the form of magnetization plateaus~\cite{onizuka-00, matsuda-13} (see Fig. 1d), the exact superlattice structures and ensuing low-energy excitations induced by magnetic fields remain unresolved. In this situation, a spectroscopic study can shed light on the field evolution of coherent magnetic quasiparticles while approaching the 1/8 magnetization plateau where Wigner-crystal-like patterns of bound triplons are created out of a diluted triplon gas, as well as on their coupling to other degrees of freedom. Raman spectroscopy is an especially valuable method to probe relevant quasiparticle excitations, as it can simultaneously detect modes corresponding to both the singlet ($\Delta S=0$) and the triplet sector ($\Delta S=\pm 1$)~\cite{fleury-68, lemmens-00, gozar-05, gozar-05b}.

Here, we present a high-field Raman scattering study on SrCu$_2$(BO$_3$)$_2$ and trace collective excitations through a correlated diluted triplon gas phase towards the 1/8 magnetization plateau of crystallized bound states. We contrast these high-field data with a high resolution, temperature-dependent Raman study of the incoherent thermal vs. quasi-coherent field-induced melting of the bosonic superstructure and the role of lattice degrees of freedom on triplon dynamics.

\section{Results}

\begin{figure}
\label{figure1}
\centering
\includegraphics[width=8cm]{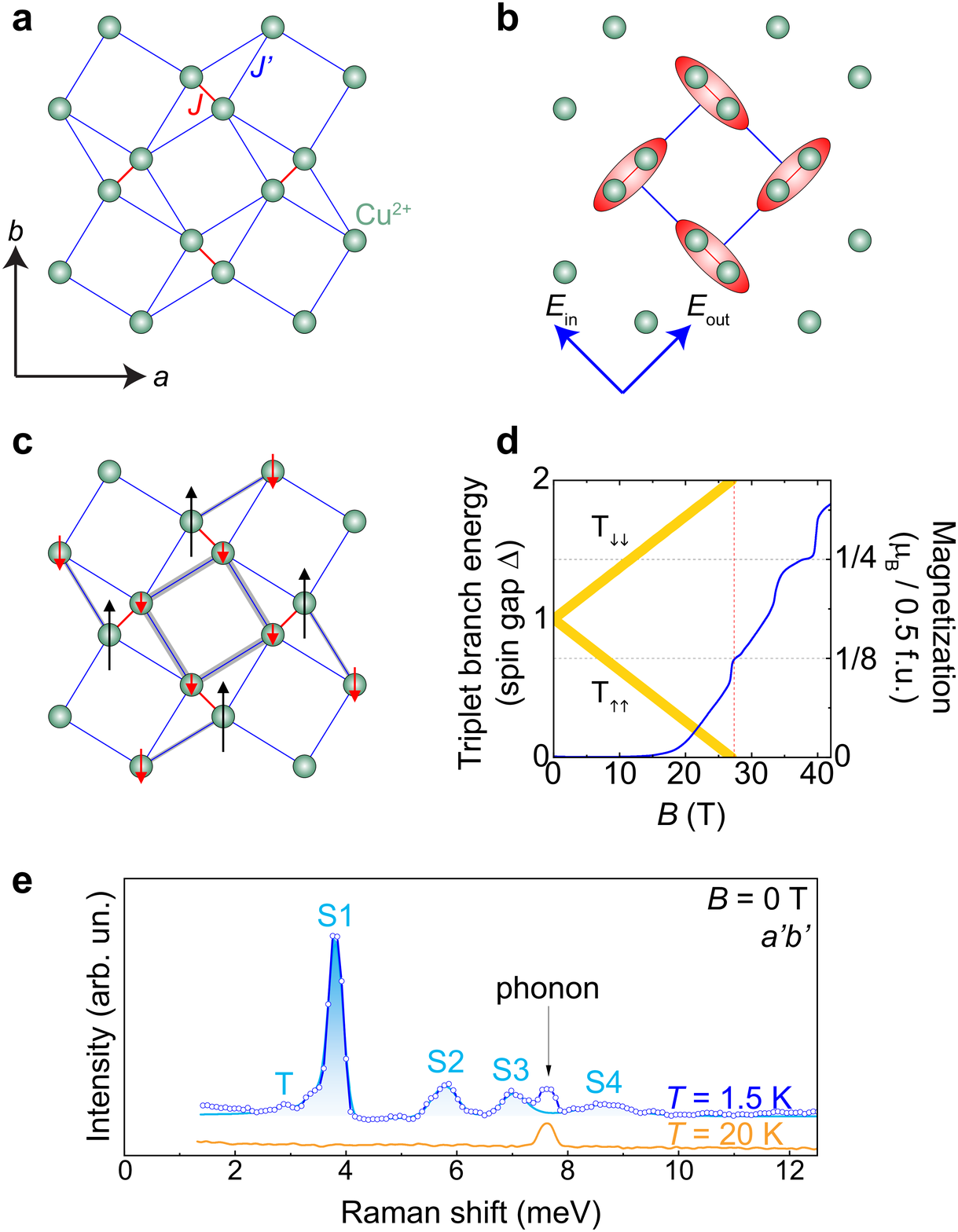}
\caption{\textbf{Magnetic ground state and exchange topology in SrCu$_2$(BO$_3$)$_2$.} \textbf{a} Magnetic exchange paths (solid red and blue lines) between individual Cu$^{2+}$ ions on the Shastry-Sutherland lattice within the crystallographic $ab$ plane. \textbf{b} Coupled orthogonal $S=0$ dimers (red ovals) result in the gapped singlet ground state at low temperatures and zero-field. The blue arrows denote the Raman scattering configuration ($a'b'$). \textbf{c} Crystallization of bound states in a pinwheel pattern at the 1/8 magnetization plateau~\cite{corboz-14}. \textbf{d} Zeeman-splitting of the triplet excitations in applied magnetic fields (yellow lines). Upon closing of the singlet-triplet gap around 27 T, the magnetization curve (blue line, measured at $T=1.4$ K with $H // c$) exhibits its 1/8 plateau (data reproduced from [\onlinecite{onizuka-00}]). \textbf{e} Zero-field Raman spectra recorded at $T = 20$ K (orange line) and $T = 1.5$ K (blue open circles) together with a fit to the rich magnetic excitation spectrum at low temperatures (a sum of Pseudo-Voigt curves; solid blue line). The individual excitations are marked as 'S' for singlet bound states and 'T' for triplet excitations.}
\end{figure}

\subsection{Magnetic excitation spectrum}

In Fig. 1e, we showcase the emergence of a rich spectrum of magnetic excitations at low temperatures (well below the spin gap $\Delta/k_\mathrm{B}=35$~K). A direct comparison between the Raman spectra taken at $T=20$~K and 1.5 K impressively demonstrates the intense and well-defined sharp excitations observed at $T=1.5$ K in the low-energy sector, which -- apart from a phonon at 7.6 meV -- are ascribed to magnetic modes. Note that all other phonons appear at higher energies ($E>15$ meV) and are energetically separated from the magnetic excitations, as we detail below. Our spectrum of magnetic excitations is in very good accordance with previous reports~\cite{lemmens-00, gozar-05, gozar-05b}. We can thus assign the observed spectral features to one singlet-triplet excitation (triplon) 'T' situated around 3 meV (i.e., its energy scale is equal to the spin gap) with $\Delta S=1$ [i.e., an excitation from the singlet ground state $\frac{1}{\sqrt{2}} \left(\lvert \uparrow \downarrow \rangle - \lvert \downarrow \uparrow \rangle \right)$ to $\lvert \uparrow \uparrow \rangle ; \lvert \downarrow \downarrow \rangle$ ] as well as to two- and higher-order singlet bound states (SBSs) 'S' with $\Delta S=0$ [e.g., $\frac{1}{\sqrt{2}} \left(\lvert \uparrow \downarrow \rangle - \lvert \downarrow \uparrow \rangle \right) \otimes \frac{1}{\sqrt{2}} \left(\lvert \uparrow \downarrow \rangle - \lvert \downarrow \uparrow \rangle \right) \to \lvert \uparrow \uparrow \rangle \otimes \lvert \downarrow \downarrow \rangle$ for S1]. We remark that the Raman-forbidden T excitation is visible due to Dzyaloshinskii-Moriya  interactions~\cite{nojiri-03}.

\begin{figure*}
\label{figure2}
\centering
\includegraphics[width=14cm]{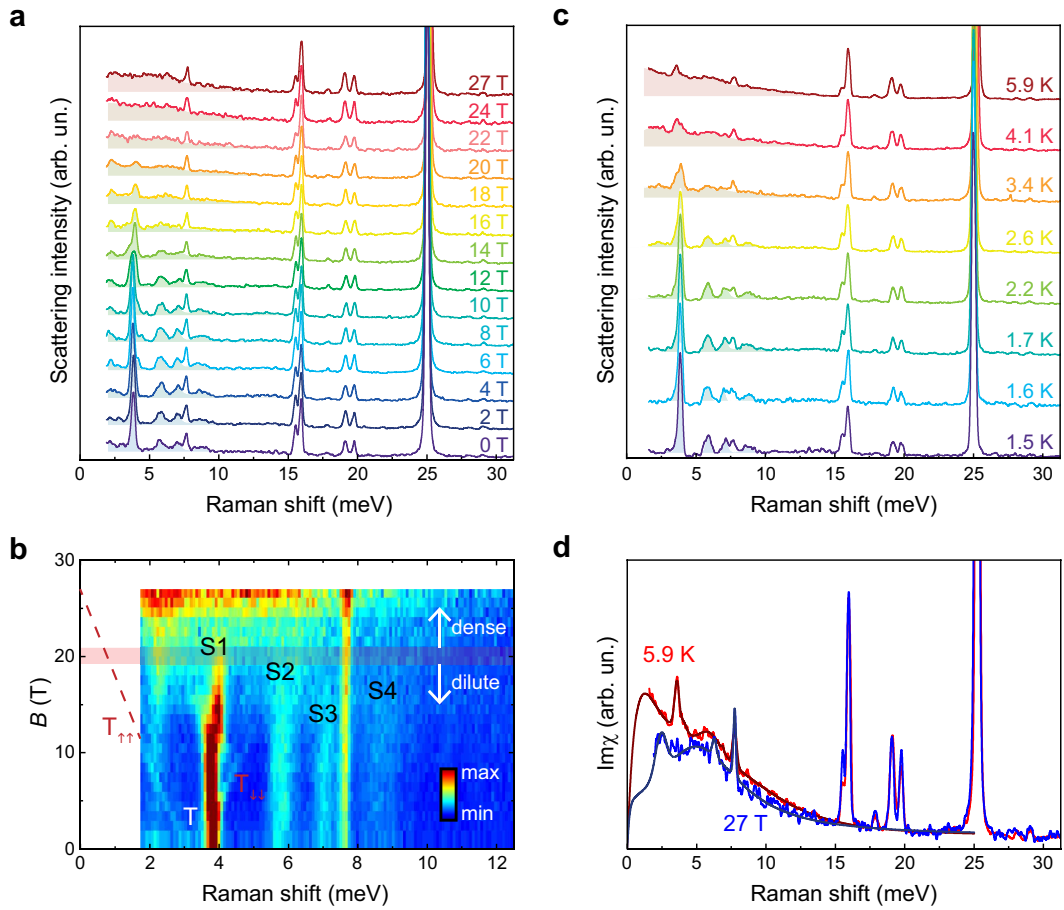}
\caption{\textbf{Field and temperature evolution of magnetic excitations.} \textbf{a} Field-dependent Raman data taken at $T = 1.5$ K and in $a'b'$ polarization. Magnetic contributions are shaded by pale colors. \textbf{b} A color contour plot of the field-dependent Raman data with an emphasis on the low-energy sector. The field-evolution of the lowest triplet excitation is traced by a dashed red line and extrapolated to $E = 0$. Dense and dilute refer to the dense magnetization plateau region and the low density triplon phase, respectively. The scale bar indicates the scattering intensity. \textbf{c} Temperature-dependent Raman data taken at $B = 0$ T and in $a'b'$ polarization. The magnetic spectral weight is shaded by pale colors. \textbf{d} A direct comparison of the Bose-corrected Raman response (Im$\chi$) measured at $B = 0$ T, $T = 5.9$ K (red curve) and at $B = 27$ T, $T = 1.5$ K (blue line). The spectral weight is extrapolated to zero energies (dark solid lines).}
\end{figure*}

To keep track of the field evolution of the collective T and S excitations, we first focus on the low-energy regime (0-10 meV) at a constant sample temperature of $T = 1.5$ K in a magnetic field range of 0 -- 27 T. The shaded backgrounds in Fig. 2a mark the spectral weight of magnetic excitations. The full data set is shown as a color-contour plot in Fig. 2b. At magnetic fields up to 10 T, we observe the Zeeman splitting of the triplet excitation 'T' into a lower branch T$_{\uparrow \uparrow}$ and an upper branch T$_{\downarrow \downarrow}$ (see the yellow lines in Fig. 1d). Extrapolating the lower branch to zero energy, we identify a closing of the singlet-triplet gap around $\Delta/k_\mathrm{B}=27$~T, in perfect agreement with the bulk magnetization and ESR data~\cite{onizuka-00,matsuda-13, nojiri-03}. At intermediate fields starting around 10 T, we notice a discontinuous hardening of the SBS S1, and a clear broadening around 12 T, coinciding with an increase in magnetization measured at the same temperature (see Fig.~1d). At higher magnetic fields around 20 T, all the SBSs S1-S4 gradually merge into a broad low-energy continuum. Although no sharp, well-defined features can be associated with this continuum, there is an enhanced spectral weight centered between 1.5 -- 3.5 meV, and a second, more shallow one between 3.5 -- 7 meV (see Fig.~2a). Based on a tensor-network ansatz, the field-induced spin configurations in SrCu$_2$(BO$_3$)$_2$ evolve with increasing fields from a region of dilute bound states into correlated states with a dense sequence of low-lying magnetization plateaus, and finally to the well-resolved 1/8 plateau~\cite{corboz-14}. Therefore, SrCu$_2$(BO$_3$)$_2$ at $B=27$ T is on the threshold to a commensurate crystallization of diluted triplet bound states, leading to sharper and better defined features.

\begin{figure*}
\label{figure3}
\centering
\includegraphics[width=13cm]{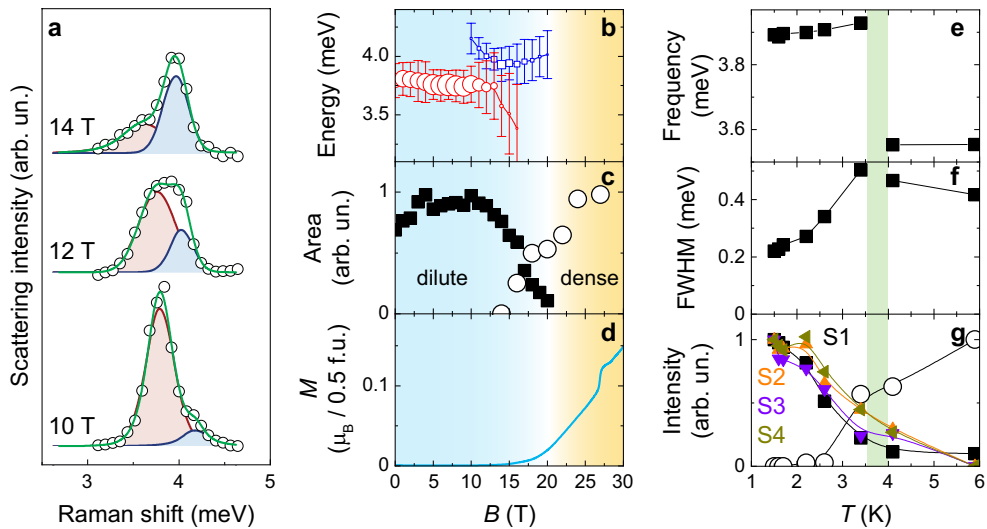}
\caption{\textbf{Field and thermal evolution of singlet bound states.} \textbf{a} Raman data measured at $T=1.5$ K and at intermediate fields of 10 T, 12 T, and 14 T (open symbols) together with fits to the individual excitations (blue and red shaded areas). The solid green lines are sums of the individual fits. \textbf{b} Field-dependence and branching of the singlet bound state S1 with increasing magnetic fields. The size of the open symbols reflects the signal intensity. Vertical bars denote the full width at half maximum. \textbf{c} Field-dependence of the integrated intensity of both contributions (black squares) together with the intensity of the broad magnetic continuum (open circles). \textbf{d} Magnetization for $B // c$ measured at $T = 1.4$ K; data reproduced from [\onlinecite{onizuka-00}]. \textbf{e} Energy and \textbf{f} linewidth of the singlet bound excitation S1 as a function of temperature. \textbf{g} Normalized intensities of singlet excitations S1--S4 as a function of temperature. The intensity of incoherent magnetic Raman scattering is shown in open circles. A crossover behavior is marked by green bars.}
\end{figure*}

We contrast these field-induced observations to the effect of thermal fluctuations  on the magnetic Raman spectrum. These experiments were carried out without applying a magnetic field. In Fig.~2c, we present the low-energy magnetic excitations (shaded by pale colors) at temperatures ranging from 1.5 K to 5.9 K. As we argue below, this temperature range roughly matches the field range 0-27 T. With increasing temperature above 5.9~K, the SBSs rapidly melt into an incoherent broad continuum due to strong scattering from thermally excited triplons~\cite{lemmens-00}. We stress that the SBSs are rapidly suppressed on the temperature scale of $k_\mathrm{B}T\approx 0.17\Delta$, amounting to only a few percent of thermally excited triplons. This signifies the thermal fragility of SBSs in the background of extremely diluted triplons and the efficient scattering of triplons on localized singlets. In contrast, in the frustrated spin ladder BiCu$_2$PO$_6$ the SBSs persist up to $k_\mathrm{B}T\approx 2\Delta$ with $\Delta/k_\mathrm{B}=32$~K and their thermal damping becomes significant above $k_\mathrm{B}T\approx \Delta$~\cite{choi-13}. The thermal fragility of SBSs in SrCu$_2$(BO$_3$)$_2$ is associated with the frustration of the two interdimer couplings. In this frustrated configuration, a thermally excited triplet polarizes six adjacent dimers of the underlying singlet state, thereby delocalizing triplons~\cite{zayed-14}. The increased kinetic energy of the thermally dressed triplons destabilizes the singlet bound states, leading to highly efficient scattering processes and to their melting into the incoherent continuum well below the characteristic temperature $\Delta/k_\mathrm{B}$. Besides, it is shown that abundant low-energy modes can lead to highly efficient scattering centers and thereby thermally destabilize any triplon configuration~\cite{honecker-16}.

A direct comparison between the $B=27$~T (at $T = 1.5$ K) and the $T=5.9$ K (at $B = 0$ T) data is shown in Fig. 2d. We observe no difference of the high-energy cut-off $\omega_{h}\approx 15$~meV between the field- and temperature-dependent magnetic continua. The partially surviving SBS around 3.6 meV at high temperatures is contrasted by the fully depleted excitation at high fields. The higher-order SBSs coalesce into the hump around 6.2 meV. For energies above 5~meV, the spectral shapes of the 27~T and $T=5.9$ K data look alike except for incoherent spin fluctuations, which are more pronounced in the $T=5.9$ K spectrum. The hint of an excitation gap at lowest energies together with an emerging mode around 2.5 meV in the high-field spectrum raises an exciting prospect of assigning a distinct spectrum of well-defined magnetic excitations to an emerging superlattice at higher fields. Although a partial thermal melting may occur even at $T=1.5$ K due to the thermal fragility of crystallized bound states, the observation of remnants of the 1/8 plateau at elevated temperatures as high as 4.2 K underlines the robustness of this particular superlattice pattern~\cite{kageyama-99}. In contrast, plateaus at higher fields such as 2/15 and 1/6, which have been reported at sub-Kelvin temperatures, appear to have melted already at 1.4 K~\cite{takigawa-13, onizuka-00}.

Having this scenario in mind, we inspect the field and thermal evolution of the low-energy magnetic excitation spectrum in detail. As we mentioned above, in an intermediate field of about 12 T, the SBS S1 exhibits a slight broadening. A close look at the data reveals the emergence of a shoulder as the origin of this broadening. As shown in Fig.~3a, the SBS S1 is decomposed into two components. We stress that the splitting of the S1 mode along with the field evolution of the T excitation is only observed for applied magnetic fields, and not at raised temperatures. With increasing magnetic field the new mode gains its intensity against the initial S1 mode. In Fig. 3b, we plot the fitting parameters of the SBSs, namely, their energy, linewidth (indicated by the vertical bars), and intensity (proportional to the symbol size) as a function of $B$. The total integrated intensity (i.e., the sum of both contributions) is nearly constant at low fields, followed by a continuous decrease above 12 T, as shown in Fig. 3c, black squares. Simultaneously, the broad background continuum gains in intensity [open circles in 3(c)]. The energy-field diagram at $B=10-16$~T is characteristic of the anti-level crossing between the two SBSs  (denoted by the green bar), originating from the competition between strongly correlated higher-field triplons forming incommensurate superlattice structures, and weakly correlated lower-field triplons. We recall that at the corresponding fields the magnetization gradually starts to increase, signaling the existence of field- and thermal-induced triplons (see Fig. 3d). In this light, the $B=10-16$ T magnetization process is dictated by the competing quantum and thermal triplons. The newly emerging SBS is observed up to 20~T, that is, in the weakly-interacting triplon regime and vanishes above 20~T, where the magnetization steeply increases. This suggests that the triplon gas phase in the dense magnetization plateau regime above 20~T is highly correlated, providing a rationale for the different low-energy behavior between the temperature and field data shown in Fig.~2d.

In Figs. 3e-g, the frequency, linewidth, and intensity of the SBSs are plotted as a function of temperature. Thermally activated pairs of triplets constituting SBSs correlate antiferromagnetically with their neighboring ones, and -- in contrast to the field evolution case -- a two-peak structure is absent over the measured temperature range. The significant demise of the modes' spectral weight and the appearance of an incoherent background continuum underlines the effective scattering processes of thermally activated SBSs. Moreover, the energy of the SBS S1 first gradually hardens with increasing temperature, followed by an abrupt softening around 3.5 K. This energy dependence is distinctly different from the field-induced generation of triplons, and instead is reminiscent of a thermally-induced unbinding of a multi-particle bound state~\cite{choi-13, wulferding-20}. This supports the notion that the temperature behavior of triplons is governed by correlated dynamics of the dressed triplons~\cite{zayed-14}. The correlated triplon dynamics provides a rationale for a rapid thermal melting of short-lived SBSs, resulting from the fact that the kinetic energy of the thermal triplons increases faster than the potential energy. In contrast, triplons that do not overlap with others may survive as long-lived SBSs up to higher temperatures. The field-induced triplons differ from the thermally populated triplons in that the Coulomb interaction between them dominates over their kinetic energy, having a strong tendency to the crystallization of triplon composites~\cite{corboz-14}.

\subsection{Field- and temperature dependence of phonon modes}

\begin{figure*}
\label{figure4}
\centering
\includegraphics[width=16cm]{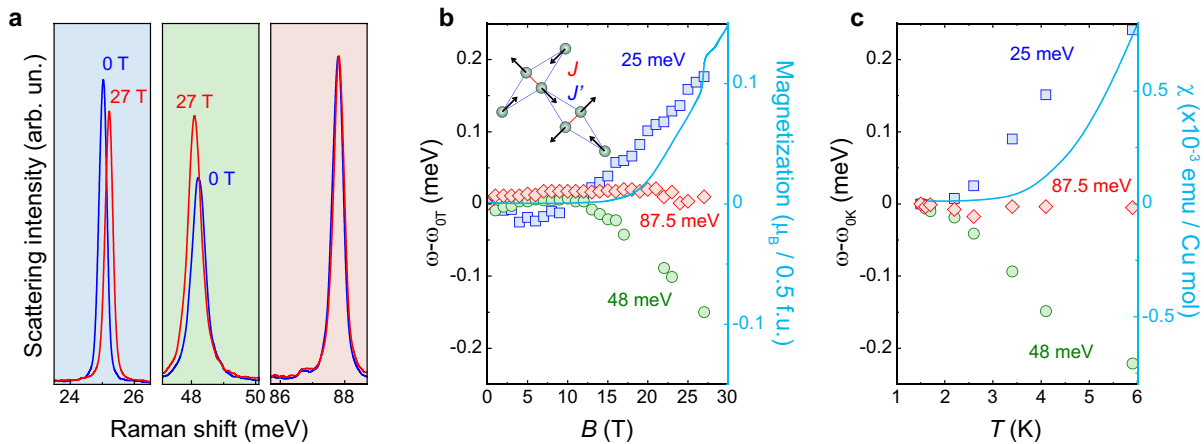}
\caption{\textbf{Field and thermal evolution of selected phonon modes.} \textbf{a} Zoom into three phonon modes at 25 meV, 48 meV, and 87.5 meV measured at 0 T (blue curve) and 27 T (red curve). \textbf{b} Normalized frequencies ($\omega-\omega_0$) of the three phonon modes as a function of an applied field, measured at $T = 1.5$ K. The blue squares correspond to the pantograph mode at 25 meV, its atomic displacement pattern is sketched in the inset. For comparison, the magnetization curve is plotted on top (blue line; data reproduced from [\onlinecite{onizuka-00}]). \textbf{c} Normalized frequencies of the three phonon modes measured at $B = 0$ T, together with the magnetic susceptibility (solid blue line; data reprinted with permission from [\onlinecite{kageyama-99}], $\textcopyright$ 1999 American Physical Society) as a function of temperature.}
\end{figure*}

Next, we address the influence of temperature and magnetic field on lattice degrees of freedom. Here, we concentrate on three phonon modes at 25 meV, 48 meV, and 87.5 meV, shown in Fig.~4a measured at two different magnetic fields and at $T=1.5$ K. While the latter phonon shows no field dependence, the former two clearly shift in energy in applied fields. In Fig.~4b, we plot the phonon energies of these three selected modes as a function of a magnetic field. The phonon at 25 meV (blue squares) corresponds to the so-called pantograph mode~\cite{radtke-15, bettler-20}. Its atomic displacement (sketched in the inset of Fig. 4b) is subject to strong spin-phonon coupling, as it directly modulates Cu-O-Cu superexchange paths on the dimer units and interdimer superexchange paths through the BO$_3$ units. As such, it can serve as a valuable local probe for detecting the alteration of magnetic interactions.

Indeed, we find that there is a close relation between the energy increase of the pantograph mode and the bulk magnetization. This demonstrates that the population of excited triplets renormalizes the energy of the pantograph mode. Similarly, the phonon at 48 meV (green circles) decreases in energy upon applying magnetic fields; its absolute change in energy with magnetic field mimics the $M-B$ curve. These experimental observations highlight the existence of substantial spin-phonon coupling in SrCu$_2$(BO$_3$)$_2$~\cite{choi-03}. They are also in agreement with earlier reports of sound-wave anomalies observed at the 1/8 magnetization plateau at finite temperatures~\cite{zherlitsyn-00}. On the other hand, a phonon at 87.5 meV remains largely unaffected by high magnetic fields and can therefore be used as an intrinsic reference. Figure~4c compares the effect of thermal heating on phonon energies of the three modes. For the phonons at 25 meV and at 48 meV we find a similar magnetic field dependence with opposite sign of the respective coupling constant, which is strikingly similar to the field-induced effects, and closely follows the $\chi(T)$-curve (plotted in blue). Note that the hardening and softening is also distinctly different from phonon anharmonicity~\cite{balkanski-83}. This further highlights spin-phonon coupling as the underlying mechanism for the anomalous temperature- and field-dependence of these two modes. Again, the phonon at 87.5 meV remains unaffected in this narrow temperature range. A comparison between field- and temperature dependence allows us to identify a similar energy scale, with $B=27$ T $\simeq$ $T=4.5$ K. We stress that both the SBSs and the pantograph phonon bear a close resemblance to their field and temperature dependence. This implies that triplon dynamics commonly underlies the thermal and field behavior of spin and lattice degrees of freedom.

\section{Discussion}

By measuring magnetic and phonon excitations as a function of field and temperature, we are able to uncover the underlying mechanism of quantum and thermal melting of triplons. To establish a correlation between them, we have chosen the temperature range of $T=1.5-5.9$~K and the field range of $B=0-27$~T, which give a comparable energy scale. The key finding is that the overall field and temperature evolution of SBSs looks similar, as inferred from Fig.~2a and 2b. Nonetheless, some discrepancy emerges at energies below 5~meV (see Fig.~2d). The similarity is due to the fact that the density of the thermal and field-induced triplons is nearly identical in the measured parameter range. On the other hand, the disparity reflects the distinct character of the thermally dressed vs the field-induced triplons.

At finite temperatures, randomly populated triplons are created out of a singlet vacuum state. Due to the orthogonal, frustrated dimer configuration, the thermally excited triplons are dressed by weakly interacting neighboring singlets. As such, the kinetic energy of the thermally excited triplons slightly increases. In contrast, application of an external field toward the 1/8 magnetization plateau renders low-lying states quasi-degenerate with the singlet ground state. In this case, the field-induced triplons are given by a superposition of the singlet and the higher-energy states. In a plateau phase with an increased number of populated triplons, the dominant Coulomb repulsion is inclined to form crystals of bound states. Given that the field-variation measurements were conducted at $T=1.5$~K, superlattice crystals -- if present -- are not in the true quantum regime. Thermal fluctuations partly melt the spin superstructure. Nonetheless, the structured spectrum below 5~meV (shown in Fig.~2d) conveys a tantalizing hint for the existence of an exotic field-induced phase such as a Wigner glass~\cite{chakravarty-93}. This conclusion is further supported by the coexistence of two SBSs in the intermediate field interval of $B=10-16$~T where a diluted triplon gas is created and the appearance of a 2.5 meV excitation in the similar field range. This is interpreted in terms of a competition between a Wigner crystallization and a binding process.

Our $T$ and $B$ comparative study of collective excitations showcases rich many-body correlated physics. The triplon density and interaction energy are delicately conditioned by temperature and field. Given that triplons and their composites lie close in energy, future spectroscopic investigations extending into the sub-Kelvin, true quantum regime will bring a profound understanding of the kinetics and dynamics of correlated bosonic quasiparticles.

\section{Methods}

\textbf{Sample growth and preparation.} Single crystals of SrCu$_2$(BO$_3$)$_2$ were synthesized by the traveling solvent floating zone method, as reported previously~\cite{kageyama-crystalgrowth-99}. Subsequently, a specimen was oriented via the Laue diffraction method and cut along the crystallographic axes to a volume of about 4 mm $\times$ 3 mm $\times$ 1 mm.

\textbf{Raman scattering experiments.} High field experiments at $T = 1.5$ K were carried out at the resistive magnet of Cell 5 at the Nijmegen High Field Magnet Laboratory in Faraday geometry, with $B // c$. Raman spectra were collected using a $\lambda = 488$ nm laser (Coherent Genesis MX) and a Horiba FHR-1000 monochromator equipped with a volume Bragg grating notch filter set and a liquid-Nitrogen cooled CCD (PyLoN eXcellon). The incident laser power was kept well below 0.1 mW to minimize heating effects. Temperature-dependent Raman data at $B = 0$ T were recorded using a $\lambda = 532$ nm laser (Torus, Laser Quantum), and the scattered light was dispersed through a Horiba T64000 triple spectrometer onto a Dilor Spectrum One CCD. All Raman scattering experiments have been carried out in crossed $a'b'$ polarization, i.e., incoming and scattered light polarized perpendicular to each other and rotated by 45$^{\circ}$ from the crystallographic $a$ and $b$ axes.

\section{Acknowledgments}
This work was supported by the Institute for Basic Science in Korea (Grant No. IBS-R009-Y3), and by HFML-RU/NWO-I, a member of the European Magnetic Field Laboratory (EMFL).  K.Y.C. was supported by the National Research Foundation (NRF) of Korea (Grants No. 2020R1A2C3012367, and No. 2020R1A5A1016518). P.L. was supported by DFG Excellence Cluster QuantumFrontiers, EXC 2123, DFG Le967/16-1, DFG-RTG 1952/1, and the Quantum- and Nano-Metrology (QUANOMET) initiative of Lower Saxony within project NL-4.

\end{document}